\documentstyle{mn}
\title[A search for hidden white dwarfs in the ROSAT EUV survey II]
{A search for hidden white dwarfs in the ROSAT EUV survey II:
Discovery of a distant DA$+$F6/7V binary system in a direction of low density
neutral hydrogen}

\author[M.\,R. Burleigh et al. ]
{M.\,R. Burleigh$^1$ and M.\,A. Barstow$^1$ and J.\,B.
Holberg$^2$\\
$^1$ Department of Physics and Astronomy, University 
of Leicester, University Rd., Leicester, LE1 7RH \\
$^2$ Lunar and Planetary Laboratory, Gould-Simpson Building, University
of Arizona, Tucson AZ 85721, USA \\
}
\date{May 27th 1998}

\def\IUE{\it IUE\rm }
\def\TD-1{\it TD-1\rm }
\def\tkev{\thinspace{ke\kern-.15em V}}
\def\tev{\thinspace{e\kern-.15em V}}

\begin{document}
\maketitle

\begin{abstract}

The ROSAT Wide Field Camera (WFC) survey of the Extreme Ultraviolet (EUV)
has provided us with evidence for the existence of a previously
unidentified sample of hot white dwarfs in unresolved, detached binary
systems. These stars are invisible at optical wavelengths due to the
close proximity of their much more luminous companions (spectral type K or
earlier). However, for companions of spectral type $\sim$A5 or later the
white dwarfs are easily visible at far-ultraviolet (far-UV) wavelengths,
and can be identified in spectra taken by IUE. Sixteen such systems have
been discovered in this way through ROSAT, EUVE and IUE observations,
including four identified by us in Paper I (Burleigh, Barstow and Fleming
1997). In this paper we report the results of our continuing search
during the final year of IUE operations. One new system, RE J0500$-$364
(DA$+$F6/7V), has been identified. This star appears to lie at a
distance of between $\sim$500$-$1000pc, making it one of the most distant white
dwarfs, if not the most distant, to be detected in the EUV surveys. 
The very low line-of-sight 
neutral hydrogen volume density to this object    
could place a lower limit on the length of the $\beta$~CMa 
interstellar tunnel of diffuse gas, which stretches away from the Local Bubble 
in a similar direction to RE J0500$-$364. 

In this paper we also analyse a number of the stars observed where no
white dwarf companion was found.  Some of these objects 
show evidence for chromospheric and
coronal activity. Finally, we present an analysis of the previously known
WD$+$active F6V binary HD27483 (B\"ohm-Vitense 1993), and show
that, at T$\approx$22,000K, the white dwarf may be contributing 
significantly to the observed EUV flux. If so, 
it is one of the coolest such stars to be detected in the EUV surveys.

\end{abstract}


\section {Introduction} 

The vast majority of the $>$2000 known white dwarfs (McCook and Sion 1998) 
are isolated stars discovered at optical wavelengths by virtue of their 
photometric colours or proper motions. In either case, there is a strong bias  
against detecting any white dwarfs in unresolved binary systems.
A companion star of type K or earlier will completely dominate
the optical spectrum of the white dwarf, effectively rendering it invisible. 
Indeed,
Sirius B, the first white dwarf to be discovered, would never have been 
resolved from the A1 dwarf 
Sirius were it not for the close proximity of the system to Earth (2.64pc).

Prior to the ROSAT and Extreme Ultraviolet Explorer (EUVE) surveys, a 
small number of unresolved white dwarf/main sequence binaries had been discovered
serendiptously. For example, 
the white dwarf in the V471 Tauri system was found as a result of an 
eclipse by its active K2V companion. In addition, a number of white dwarfs have been 
accidently discovered during various  far-ultraviolet (far-UV) observations 
of the normal stellar companions by the International Ultraviolet Explorer (IUE),
e.g. $\zeta$ Cap (B\"ohm-Vitense 1980), 56 Peg (Schindler et al. 1982) and 4
$\sigma$$^1$ Ori (Johnson and Ake 1986), although Shipman and Geczi (1989)
systematically studied the then existing IUE archive for white dwarf companions to G, K
and M stars, and found no further examples. 

Now, however, 
the extreme ultraviolet (EUV) surveys of the ROSAT Wide Field Camera 
(WFC, Pounds et
al. 1993, Pye et al. 1995) and  EUVE (Bowyer et al. 1994, 1996) have 
provided evidence for the existence of a substantial sample of these previously
unknown white dwarfs, through the detection of EUV radiation.  
Sixteen new systems have been discovered in this way, including $\beta$
Crateris (A2IV$+$WD, Fleming et al. 1991), KW Aur C (F4V$+$DA, Hodgkin et al.
1993), HD18131 (K0IV$+$DA, Vennes et al. 1995), 
RE J0357$+$283 (K2V$+$DA Jeffries, Burleigh and Robb 1996), and the
latest two, RE J0702$+$129 (K0Ve$+$DA, Vennes et al. 1997), and HR2875 
(B5V$+$WD, Vennes, Berghoefer and Christian 1997, Burleigh and Barstow 1998). 
Detailed studies of groups of these objects have been undertaken by Barstow et al.
(1994) and Burleigh, Barstow and Fleming (1997, paper I).  Positive 
identifications have been made in each case through follow-up observations in the
far-UV with IUE, since for companions later than $\sim$A5 the hot white dwarf is easily
visible at these wavelengths (the companion to the B5V star HR2875 was
identified through an EUVE spectrum, since the B star still dominates the
spectrum at far-UV wavelengths). 

It is well established that over half of all stars are members of binary 
or multiple systems, yet the overwhelming majority of catalogued white 
dwarfs are isolated objects. 
The new population of optically hidden white dwarfs 
emerging from the EUV surveys has profound implications, 
therefore, for our knowledge of the white dwarf luminosity function, space density and
formation rate (e.g. Fleming, Liebert and Green 1986). Observations of white dwarfs in
detached binary systems also allow us to place constraints on their 
evolutionary models (e.g. de Kool and Ritter 1993).

In this paper we report the results of a continuing search for more of these 
binaries in
the ROSAT WFC catalogue, during the final year of IUE operations. We have
discovered one new unresolved white dwarf$+$main sequence star binary (RE
J0500$-$362)\footnote{Originally reported, with a preliminary analysis, 
in the proceedings of the 1996 European White Dwarf Workshop 
by Burleigh and Barstow (1997).}. 
In the cases where no white dwarf was detected, a number
of the target stars show evidence for chromospheric activity. We also present an
analysis of the previously known white dwarf$+$active star system HD27483
(B\"ohm-Vitense 1993). We show that the white dwarf is contributing to
the EUV flux and should be included in the growing list of EUV-emitting 
optically-hidden hot white dwarfs in binary systems.

\section{Detection of the sources in the ROSAT WFC EUV survey}
 
The ROSAT WFC EUV all-sky survey was conducted between July 1990 and 
January 1991. Two broad band filters were utilised (designated S1 and S2), 
and most of the count rates quoted in this paper (see Table 1) 
are taken from the 2RE 
catalogue (Pye et al. 1995). This revised list contains 479 EUV sources, as 
compared with 383 in the original Bright Source Catalogue (Pounds et al. 
1993). The 2RE catalogue was compiled from the original survey data with 
improved methods for source detection, background screening etc. The 
resulting count rates are equivalent to on-axis, at-launch values.
 
The ROSAT PSPC X-ray survey was conducted simultaneously with the WFC. 
The soft (0.1$-$0.4keV) and hard (0.4-2.4keV) band count rates (Table 1) 
were obtained via the World Wide Web from 
the on-line All-Sky Survey Bright Source Catalogue, maintained by the 
Max-Planck Institute in Germany (Voges et al. 1996). 
All of the X-ray flux from hot white dwarfs is expected to lie within 
the soft band.
 
The EUVE all-sky survey was conducted in four pass bands between July 1992 and 
January 1993. These count rates are also given in Table 1, and are mainly taken 
from the Second EUVE Source Catalog (Bowyer et al. 1996) which, like the 
ROSAT WFC 2RE catalogue, includes better source detection algorithms and 
improved reliability. 
 
\subsection{Selection of candidate hidden white dwarfs}
 
Hot white dwarfs in unresolved binaries with companions of spectral type 
K or earlier are virtually impossible to discern from optical spectra alone. 
However, it is possible to unambiguously identify these hidden white
dwarfs 
in far-UV spectra taken by IUE. The problem is how to select likely 
candidates from just their EUV and soft X-ray count rates.
 
Most of the hot white dwarfs discovered by ROSAT have very soft spectra 
compared to normal stars, particularly where the interstellar hydrogen 
column density is low. The ratio of the WFC S2 to S1 count rates can often 
exceed a factor of two. Additionally, no photons are usually detected from a
white 
dwarf above the 0.28keV carbon edge of the ROSAT PSPC. All other X-ray and 
EUV-emitting astronomical objects generally have spectra extending to
higher energies. Thus, many of the stars originally selected for 
observation by IUE were relatively bright 
EUV sources with very similar colours to known hot white dwarfs (e.g. 
KW Aur C, Hodgkin et al. 1993). The success rates of the early searches
by, for example, Barstow et al. (1994) were very high, and appeared to
represent the tip of an iceberg. 
 
However, many of the white dwarfs in the ROSAT WFC survey are relatively 
faint EUV sources, indistinguishable by count rate ratios alone from 
coronally active objects. Further selection criteria, 
in addition to simple EUV colour and brightness, needed to be applied. 
In Paper I (Burleigh, Barstow and Fleming 1997) 
we conducted a search for fainter, less obvious examples of  
these binary systems, with a $\sim$40\% success rate. Candidates were selected, 
in particular, from normal stars that had been observed in the WFC optical 
identification programme (Mason et al. 1995), and where little or no 
evidence of activity had been found. In some cases, e.g. HD2133 
(RE J0024$-$741), a hidden white dwarf was indeed detected in an IUE SWP 
spectrum. In others, e.g. HR2468 (RE J0637$-$613), the IUE observations 
revealed evidence of chromospheric and coronal emission that had eluded 
the optical identification team. 
 
In this paper we report the results of a further search for hidden 
white dwarfs during the last year of IUE operations (1995/96). Thirteen 
candidates were selected and observed, including many that were detected 
for the first time in the reprocessed ROSAT WFC 2RE data. These are, 
in general, faint EUV sources. In most cases, the target stars were not 
known to be active, and had S2/S1 count rate ratios $>$2. In a separate 
optical programme, a number of the unidentified 2RE fields were observed 
to try to determine the counterpart to the EUV source. In some cases, e.g. 
RE J0500$-$364, 2RE J0222$+$503 and 2RE J0232$-$025, only one  relatively
faint star (V$>$10) was visible in the field. Although a few EUV sources
(mainly red dwarfs) are active stars with V$>$10 (e.g. Proxima Cen, 
M5Ve, V$=$11.1), most are much brighter than this. Thus, for these
sources, 
a hidden hot white dwarf companion was a feasible explanation for the EUV 
radiation, and they were added to the IUE target list.
 
\subsection{UV spectroscopy}
 
A log of all the far-UV observations is given in Table 2. The white dwarf 
companion to HD27483 was serendipitously discovered in October 1992 by 
B\"ohm-Vitense (1993), independently of the ROSAT survey, during an 
IUE survey of Hyades F stars. RE J0500$-$364 was first observed in a  
short half-hour exposure in November 1995 (SWP56217), 
and a faint white dwarf was 
detected. A follow up spectrum, with an 8 hour exposure to achieve the
required signal-to-noise, was obtained a month later. 
 
The  programme was unfortunately 
cut short in February 1996 when a gyro failed 
on IUE, limiting observations to objects for which a 
bright (V$<$8) guide star was available. This was not the case for any of 
our remaining targets. 
 
Two spectra were affected by the so-called `159DN anomaly' (see Table 2). 
A number of pixels were 
assigned the flux value 159DN when they should in fact have had a higher 
value. However, on close inspection it was found that only a small fraction
of each of these two spectra was contaminated, and their scientific  
usefulness was not compromised.

\section {IUE Data Reduction}

All of the spectra have been 
processed with NEWSIPS (New Spectral Image Processing System). 
NEWSIPS spectra contain a number of significant geometrical and photometric 
corrections which enhance spectral signal-to-noise and 
improve the photometric reliability of the data (e.g. a correction for the 
degradation of the detector with time, Bohlin and Grillmair 1988).
Recent analysis by Garhart (1997)  
shows that, since 1993, the  SWP camera sensitivity has in fact     
degraded at a rate greater than predicted, so that the current 
NEWSIPS calibration underestimates the SWP fluxes by up to 5\%. 
Archival IUE data from 1993-96 are now being reprocessed 
to include a new degradation correction. However, this data was not
available to us when we were preparing this paper, 
although we would expect the effect of the new calibration to
be  relatively minimal on the determination of any white dwarf
parameters (for example, the errors on the flux values for the RE
J0500$-$364 IUE SWP data are typically of order $\sim$10\% anyway).

\section {Analysis}

The method used to analyse the far-UV and EUV data for the hidden white dwarfs and 
the active stars detected and observed during this search has been discussed in detail
in Paper I (Burleigh, Barstow and Fleming 1997). A summary is given here. 

\subsection{Hidden white dwarfs - far-UV data}

In binary systems like these, it is not possible to use the H 
Balmer line profiles to measure 
temperature and gravity. However, the IUE SWP spectra can  
be used to estimate these parameters 
by matching the observed Lyman $\alpha$ profile and the UV
continuum (the region uncontaminated by the companion) with synthetic spectra. 
Unfortunately, it is not possible to get an unambiguous determination of T and log g 
from a single spectral line. Instead, a range of possible models is determined by
stepping through values of the surface gravity, from log g$=$7.0 to 9.0, 
and finding the best fit temperature and normalisation at each point. 

We compare the observed far-UV data with fully line blanketed,  
homogeneously mixed H$+$He, LTE model atmospheres, spanning a temperature range from 20,000K to
100,000K, and kindly supplied by Detlev Koester (e.g. Koester 1991). 
In this analysis, we assume the white dwarfs have pure hydrogen envelopes. 
The spectral fitting is
conducted with the XSPEC programme (Schafer et al. 1991), which calculates a
chi-squared statistic for the fit between the data and the model, and which is then
minimised by incremental steps in the free parameters. There is no need to take into
account any interstellar component in the analysis of the Lyman $\alpha$ profile, because
for columns greater than a few $\times$10$^{19}$ the white dwarf is unlikely to be
detected in EUV surveys. 

The white dwarf radius and mass are calculated using the 
temperature-dependent evolutionary models 
of Wood (1995), which assume thick (10$^{-4}$M$_\odot$) H layers, 
and the radii are then used to estimate the distance from 
the model normalisation parameter (which is actually the solid angle of 
the star, equivalent to [radius/distance]$^2$). 
The distances to the primaries in each case 
can be calculated from the spectral type and V magnitude, and are given in 
Table 3. The range of white dwarf temperatures and gravities that give the best
match to the distance of the primary can then be estimated. 
The V magnitude of each white dwarf is estimated from the model 
flux at 5500{\AA}, and given in Table 5.
 
\subsection{Hidden white dwarfs - ROSAT data}

Once the temperature and gravity of each white dwarf has been estimated, the 
ROSAT EUV and soft X-ray fluxes can give an indication of the level of 
photospheric opacity in the stellar atmosphere, by comparing them with 
predicted values for a pure H model. The ROSAT data is fitted 
independently from the IUE data since contamination from elements heavier 
than H and He only has an effect at EUV and soft X-ray wavelengths. We fit 
the data from the two WFC filters, and the integrated count rate in the 
0.1$-$0.28 keV PSPC band, within which all the white dwarf soft X-ray flux 
is expected to lie. It is possible that some EUV and X-ray emission 
might also originate from the normal stars in each system. 
Any detected PSPC flux at energies above 0.4 keV is an indication of the  
presence of an active companion (Table 6).

Once again we utilise Detlev Koester's fully line blanketed H$+$He models. 
These assume a homogenous distribution of hydrogen and helium, under LTE 
conditions, in the range $-$8$<$logHe/H$<$$-$3. The temperature, gravity 
and normalisation, estimated from the fit to the IUE data, are  
frozen during the modelling, but the He/H ratio is allowed to vary. The 
interstellar HI column density is also estimated by assuming that the 
local ISM is not highly ionised (i.e. there is minimal HeII absorption) and that 
the HeI/H ratio is cosmic (0.1).

The fitting is again conducted using the XSPEC programme. We consider a 
good fit to the data to correspond to the probability that a particular value 
of the reduced $\chi$$^2$ ($\chi$$_r$$^2$) can occur by chance to be 0.1 or  
greater (i.e. 90\% confidence), and a bad fit 0.01 or less (99\% confidence). 
The fits in between might not be very good, but cannot be ruled out with 
high confidence. In this analysis a good fit requires $\chi$$_r$$^2$$<$2.71, 
but until it exceeds 6.63 a model cannot be excluded with any certainty. 
Consequently, we list all model fits to the data for which $\chi$$_r$$^2$$<$6.63 
(Table 6).

\subsection{Non-detections and active stars}

A number of the stars observed with IUE where no white dwarf was detected 
are probably coronally and chromospherically active. Some of these stars 
are also comparatively 
hard X-ray sources (i.e. they are detected in the 0.4$-$2.4 keV 
band of the ROSAT PSPC), and in some of the IUE LWP spectra chromospheric MgII 
emission was seen at 2798{\AA}. 

In these cases, estimates of the L$_{EUV}$/L$_{bol}$ and L$_x$/L$_{bol}$ 
ratios have been made as an indicator of the level of activity (Table 7).  
We adopt the methods 
outlined by Jeffries (1995) and Fleming et al. (1995) to estimate these 
parameters. 

Where chromospheric MgII emission was seen the line fluxes above the
continuum level were measured
using a simple Gaussian profile, fitted to each line, after the continuum
had first been subtracted (the continuum flux was represented by a low
degree polynomial). Note that in each case the MgII emission actually
fills-in an underlying absorption dip. However, it is impossible to
estimate the depth of this absorption line, and thus the measurements
presented here are only lower limits. The measured MgII fluxes are listed
in Table 7, and have been used to estimate L$_{MgII}$/L$_{bol}$.   

\section {Discussion}

\subsection{Hidden white dwarf binaries}

\subsubsection{RE J0500$-$364}

The field of this EUV source (see Figure 1) 
was surveyed by Mason et al. (1995) during the WFC
optical identification programme. No evidence of activity was found in
the cores of Ca II H \& K in the spectrum of a $\sim$13th magnitude
star located at the centre of the source error box. Various stars were also
examined outside the error box, but none could plausibly account for the 
EUV emission. The field was also examined on 1995 September 29/30 with the
2.3m Steward Observatory telescope on Kitt Peak, as part of a programme to
search for the optical counterparts to unidentified sources in the 
ROSAT WFC 2RE catalogue. A spectrum (Figure 2) of the central star in the source
error box was obtained with the Boller and Chivens Spectrometer and
800$\times$1200 blue sensitive CCD. A 2.5" slit and 600 lines/mm grating
blazed at 3658{\AA} were used, and the data were reduced with standard
IRAF routines. Again, there was no evidence for activity. 
One untested proposition was  
that this central object could be hiding a  hot 
white dwarf companion. Therefore, the  star was added to our IUE target
list for observations during 1995/96, and a faint, $\approx$18th
magnitude hot white dwarf companion was discovered (Figure 3). 

By comparing the relative line strengths and widths 
of the primary's optical spectrum with spectra in the atlas of  
Jacoby, Hunter and Christian  
(1984), we conclude that it most closely resembles an F6/7V. It should be  
noted that the spectrum appears to be deficient in flux at the blue end, below 
$\sim$4500{\AA}. As the star was observed at a relatively high air mass 
(sec Z$=$2.94), this deficiency in counts was probably caused by differential 
atmospheric dispersion. The spectral identification is strengthened
when the IUE LWP spectrum (LWP31729, Figure 4) is compared to spectra in the IUE
Spectral Atlas (Wu et al. 1991). The data are very noisy, and have been
binned up as a consequence, but  
most closely match stars in the range F5V$-$G0V. There also appears
to be a slight excess of flux in the IUE SWP spectrum, at
wavelengths $>$1850{\AA}, above the level expected from the white dwarf 
(Figure 3). If the excess is real, then it must attributable to the primary. 
As $\sim$13th magnitude mid-late G and  K stars are not bright enough to 
be detected by IUE in this region of the spectrum, 
this also indicates that the companion is probably a late F. Taking the Guide
Star Catalogue magnitude of V$=$13.29, the system 
lies between 755 (F7V) and 870 (F6V) parsecs away, although if we
consider possible errors on the GSC magnitude ($\pm$0.3 mags), and that
the absolute magnitudes 
may be in error by up to $\pm$0.5 magnitudes, the system could be as
close as 520 parsecs or as distant as 1250 parsecs.

Is this a true binary or a chance alignment? Given that the source of the
EUV radiation is almost certainly the hot white dwarf, we can calculate the
probability of a random 13th magnitude star also falling within the IUE
aperture. According to Allen (1973), the number of stars per square degree
at the Galactic latitude of RE J0500$-$364, brighter than V$=$13, is 87.1.
The IUE large aperture is a 10"$\times$20" oval, hence we
calculate a probability of $\approx$1/250 of a chance alignment. (Note
that the point spread function of the stellar image of the 13th magnitude
star on the Digitised Sky Survey plate is $\sim$15" in radius, see Figure
1).

Assuming this is indeed a true binary, then we believe 
RE J0500$-$364 to be one of the most distant white dwarfs, if not the
most distant, to be identified in the EUV
surveys (for a comparison, see the distance estimates for the white
dwarfs detected by the ROSAT PSPC by Fleming et al. 1996, and those
detected by EUVE by Vennes et al. 1997b). 

A model fit to the IUE spectrum 
at log g$=$7.25 gives T$=$38,260K and M$=$0.40M$_\odot$,
corresponding to a distance of 830 parsecs. A fit at log g$=$8.0
corresponds to the minimum distance estimate to the system, but gives a
higher mass, M$=$0.68M$_\odot$ and higher temperature, T$=$45,000K (and
brings the star into the regime where its atmosphere may be
contaminated by elements heavier then He). The
inferred V magnitude from these models is 17.9, making this one of the
faintest white dwarfs to be detected by ROSAT. The faintest, RE
J0616$-$649 (V$=$18.4), is a rare magnetic DA (Jordan 1997), 
but there are few other hot white dwarfs detected fainter than 17th magnitude.

This source is not included in the ROSAT PSPC All-Sky Survey Bright
Source Catalogue (Voges et al. 1996), but it has
been observed and detected in X-rays during a pointed observation 
by the ROSAT High Resolution Imager
(HRI). The target was observed as part of programme to identify possible
isolated neutron star candidates in the ROSAT surveys (PI Wang). Using
the Point Source Search programme (PSS) within the STARLINK Asterix X-ray
analysis package (see Pye et al. 1995) to
analyse the HRI image, the
position of the X-ray source was found to be coincident with the WFC and EUVE
detections (Figure 1), and has a flux of 27.4$\pm$3.9 counts per 1000
sec$^{-1}$. The source shows no evidence for time variability. 

Although the HRI has
very limited spectral response (the spectral properties of the HRI are known
to vary with detector position and time in an ill-defined manner), it is
possible to obtain a crude hardness ratio. In this case, the HRI data can
then be used to test whether there is any hard X-ray emission from the F6/7V 
companion star, since no emission is expected from the white dwarf
photosphere above 0.4 keV. Figure 5 shows the HRI data plotted as a
function of pulse height distribution. The exact energy of each channel
varies temporarily and spatially, but channels 1 to 5 are roughly equivalent to
the soft (0.1$-$0.4 keV) PSPC band, and channels 6$-$16 are equivalent to
the hard (0.4$-$2.4 keV) band. As with the PSPC, a hardness ratio can be
defined by (Hard$-$Soft)/(Hard$+$Soft). In this case, the hardness ratio
$=$$-$1.0 confirming, as can be readily seen in Figure 5, that this is
a very soft source. We conclude that all of the EUV and soft X-ray emission
originates from the hot white dwarf.

Unfortunately, there is no reliable detector matrix for the HRI, and it
did not prove possible to use the X-ray count rate in the subsequent
analysis.
Therefore, since this left only the two ROSAT WFC data points to try to 
constrain the interstellar hydrogen column density and white dwarf
atmospheric parameters, 
we assumed that the white dwarf photosphere was essentially pure hydrogen 
(a reasonable assumption for T$<$40,000K, and for many hot DAs T$<$50,000K), 
and allowed only the neutral hydrogen column density to vary in the 
subsequent fitting of the EUV data. We found that the WFC S2
photometric data point is well matched by a model fit 
at log g$=$7.25, although the 
S1 flux was predicted to be slightly lower than observed. 
The hydrogen column density  given by this model is 
N$_H$$=$7.5$\times$10$^{18}$ atoms cm$^{-2}$. 
At 830 parsecs distance, this translates into a neutral hydrogen line-of-sight 
volume density of only 0.0029 atoms
cm$^{-3}$, well below the average local volume density within $\sim$80pc
of the Sun of 0.05 atoms cm$^{-3}$ (Warwick et al. 1993).
If the system is closer (530 parsecs) 
and the log g$=$8.0 model is applied, then the H
column density is higher (N$_H$$=$2.4$\times$10$^{19}$ atoms cm$^{-2}$)
and the line-of-sight volume density is also higher at 0.0146 atoms
cm$^{-3}$. Again, though, this is  well below the average local volume
density. 

Notably, this system (galactic co-ordinates l$=$239.6$^\circ$,
b$=$$-$37.2$^\circ$) lies in a similar direction to the
known exceptionally low column densities towards the B1 giant $\beta$ CMa 
(l$=$226$^\circ$, b$=$$-$14$^\circ$, Welsh 1991), and in particular
to two other ROSAT-discovered hot white dwarfs, RE J0457$-$281 and RE
J0503$-$289 (l$=$229.3$^\circ$, b$=$$-$36.2$^\circ$, Barstow et al.
1993). 

At $\approx$200pc distance, 
$\beta$ CMa is known to exist in a rareified ``interstellar tunnel''
of very low neutral gas density, which is itself an extension of the
region surrounding the Sun called the Local Bubble (Welsh 1991). 
The features of the
Bubble were first mapped out by Frisch and York (1983), and Welsh (1991)
speculates that the $\beta$ CMa tunnel may extend for {\it at least} 
300pc in 
that direction away from the Sun. Welsh (1991) also estimates the tunnel
to be about 50pc in diameter; the white dwarfs RE J0457$-$281 and RE
J0503$-$289 (which are much closer at $\approx$90pc) 
may then possibly exist in a southward extension of this tunnel, or 
more likely lie in the foreground, within the Local Bubble itself. 
 
Table 8 details the known column densities and volume densities to
these three stars, and also to $\epsilon$~CMa (B2 II, l$=$239.8$^\circ$,
b$=$$-$11.3$^\circ$), the brightest EUV source in the
sky (Cassinelli et al. 1995), which lies in a similar direction. The
line-of-sight neutral hydrogen volume density we measure for RE0500$-$364 
compares favourably with the average volume density 
to these four stars (0.0030 atoms cm$^{-3}$).
This suggets that 
RE J0500$-$364  might also lie within the $\beta$~CMa extension of the  
Local Bubble. 

If RE J0500$-$364 really does lie as far as $\sim$500 or more  
parsecs away, then it  
presents a possible lower limit to the size of any neutral gas-free corridor 
stretching away from the Local Bubble in that direction. 
The region is bounded on three sides by several OB
associations: the Orion nebula (450pc away), the CMa OB1 association
(800pc distant), and the Gum Nebula (290pc away). 
Welsh (1991) hypothesises that this tunnel may have been evacuated by a
number of supernova explosions in the last few $\times$10$^5$ years. The
injection of driven, heated, rarefied gas into an older ($\sim$10$^7$
years old) low density cavity in the local interstellar medium would
produce the large region of very low density neutral gas that we now see.

\subsubsection{HD27483}

The Hyades system HD27483 consists of two active F6V stars orbiting each other
with a period of 3.05 days. 
Although an EUV source was detected originating from the direction of this 
system in the ROSAT WFC all sky 
survey in 1990, the hot white dwarf component was identified
independently and serendipitously by B\"ohm-Vitense (1993) during an IUE SWP
observation as part of a survey of Hyades F stars (B\"ohm-Vitense 1995). 
B\"ohm-Vitense (1993) 
derived atmospheric parameters for the white dwarf of 23000$\pm$1000K and
M$=$0.6M$_\odot$. However, in the analysis of the white dwarf spectrum 
the author 
utilised the unblanketed models of Wesemael et al. (1980), assuming log
g$=$8.0, to fit the far-UV spectrum at just two points. Since
the white dwarf might be hot enough to be contributing to the observed 
EUV flux,  it could be argued that the system should be included on the
list of EUV-detected hidden white dwarfs, and thus we have decided to
re-analyse the far-UV and ROSAT data here, in the same manner as the other 
recently discovered systems.

The ROSAT WFC source RE J0420$+$138, associated with HD27483, 
was listed in the original Bright
Source Catalogue (Pounds et al. 1993), with a count rate of 15.0$\pm$4.0 counts
ksec$^{-1}$ (it was not detected in S2),  
but was not subsequently detected in the reprocessed 2RE survey
(Pye et al. 1995). The significance of the S1 detection in the 2RE survey was
4.2; sources had to exceed a significance of 5.5 in a combination of both
bands to be included in the catalogue.  
Even so, at T$\approx$23,000K the white dwarf may be contributing
to this small EUV flux, despite the fact that the two F star companions
are known to be active themselves. 

There is significant contamination in the IUE SWP spectrum (see Figure 6) 
at the long wavelength end from the two F6V
star companions, but this falls to zero by  1600{\AA}. Therefore, we were
able to use the continuum flux up to this point. B\"ohm-Vitense 
estimated a distance of 47.6 parsecs to this system. We can further 
constrain this figure with the recently published {\it Hipparcos} parallaxes
(ESA 1997), where the measured value for HD27483 is 21.8$\pm$0.85mas, 
corresponding to a distance of 45.9$+$1.80$/$$-$1.75 parsecs. 
The spectral model which best matches this distance has log g$=$8.5 and   
T$=$22,000, and the corresponding stellar mass is 
M$=$0.94M$_\odot$. The white dwarf age is then 1.4$\times$10$^8$ years, 
in comparison with the age of the Hyades
cluster, $\sim$7$\times$10$^8$ years (B\"ohm-Vitense 1993). Given that
the white dwarf might have a higher mass than is the average for these
stars ($\approx$0.6M$_\odot$, Marsh et al. 1997), we can 
estimate its progenitor mass and place a possible lower limit on the
maximum mass for white dwarf progenitor stars in the Hyades cluster. 
From Wood (1992): 

\vspace{0.35cm}

$M_{\em WD}$$=$Aexp(B$\times$$M_{\em MS}$) 

\vspace{0.35cm}

where~A$=$0.49462M$_\odot$ and B$=$0.09468M$_\odot$$^{-1}$. We find, 
for $M_{\em WD}$$=$0.94M$_\odot$, $M_{\em MS}$$=$6.7M$_\odot$. 

The main sequence lifetime of a 6.7M$_\odot$ star is in fact  
significantly shorter than 560 million years (the difference between the
Hyades age and the white dwarf cooling age, e.g. Schaller et al. 1992). 
This suggests that the white dwarf is, in reality, probably lower in mass than
0.94M$_\odot$. However, in order to unambiguously and tightly constrain 
the fundamental parameters of this star, we will need
to obtain a spectrum of the H Lyman series with an instrument such as the
forthcoming {\it FUSE} mission (see also Section 6).

The WFC source is coincident with a ROSAT PSPC X-ray source, with a
total count rate of 130.6$\pm$19.8 counts ksec$^{-1}$, including a
detection in the upper band. This confirms that at least one of the two
F6V companion stars is active, as the hard X-rays could not have
originated from the white dwarf. The ROSAT data points cannot be
matched with any of the white dwarf models (which assume a homogeneous
atmospheric mixture of H and He). This implies that the
active star(s) must be providing a significant fraction of  
the EUV flux, since little or 
no heavy element contamination is expected in the white dwarf photosphere in 
this cool temperature regime. It is even possible that
there is no flux at all from the white dwarf at these wavelengths. 
The contribution of the white dwarf to the S1 count rate can, however, 
be estimated. Another hot WD$+$MS binary, V471
Tauri,  is detected by ROSAT in the Hyades cluster (Barstow et al. 1992).
After subtracting the contribution from
the active K2V companion, Marsh et al. (1997) use the WFC count rates to  
estimate the H column density to this system  
(8.52$\times$10$^{18}$ atoms cm$^{-2}$). Adopting the same column
density to HD27483,  
assuming a pure H atmosphere, and using the parameters derived
from the log g$=$8.5 model, the white dwarf is predicted to 
contribute 5.4 counts
ksec$^{-1}$ to the S1 flux (i.e. $\approx$1/3 of the 15 counts ksec$^{-1}$
detected). How is this count rate affected by
uncertainties in the H column density? In fact, from EUVE
spectra, Dupuis et al. (1997) derived a much
lower H column density to V471 Tauri of 1.5$\times$10$^{18}$. 
Using this value, in the log
g$=$8.5 model, the white dwarf  contributes 7.6 counts
ksec$^{-1}$ to the  S1 flux (i.e. $\approx$1/2 of the observed flux). 
This hot degenerate companion to 
HD27483 could itself, then, be regarded as a real EUV source. At
T$\approx$22,000K, this would make it one of the coolest white dwarfs  
to be detected in the EUV surveys.

The combined X-ray luminosity of the two F6V stars in the HD27483 system can also be
estimated, by subtracting the contribution of the white dwarf to the ROSAT 
PSPC lower band flux. The PSPC count rates are   
98$\pm$14 counts ksec$^{-1}$ in the softer 0.1$-$0.4keV band, and 33$\pm$9 counts
ksec$^{-1}$ in the harder 0.4$-$2.4keV band (Voges et al. 1996). 
In the log g$=$8.5 model, assuming a column density of
8.52$\times$10$^{18}$ atoms cm$^{-2}$, the white dwarf flux in the
0.1$-$0.4keV band is found to be 36.0 counts ksec$^{-1}$. Eliminating  
this from the total PSPC lower band rate and following the
method detailed by Fleming et al. (1995), we find L$_x$$=$1.7$\times$10$^{29}$
ergs sec$^{-1}$, and L$_x$/L$_{bol}$$=$7.8$\times$10$^{-6}$.  

\subsection{Non-detections and active stars}

\subsubsection {BD$+$49$^\circ$646 (2RE J0222$+$50)}

This unclassified star was only observed by the SWP camera (SWP56333), and
there was no flux visible above the background. If
BD$+$49$^\circ$646 is a G or K star, then we probably would not 
detect it in this waveband anyway. No emission features are visible in
the UV spectrum, but the EUV source is coincident with a ROSAT
PSPC hard X-ray source, and the possibility must remain that 
BD$+$49$^\circ$646 is coronally  active.  Alternatively, the EUV/X-ray 
source might be another object in the field, or it is possible that
the target was missed altogether in the IUE observation.

\subsubsection {2RE J0232$-$02}

As with the WD$+$MS binary RE J0500$-$364 (discussed above), the
field of this WFC source was originally observed in 1995 with 
the 2.3m Steward Observatory telescope at Kitt Peak, as part of a programme to try to 
identify the remaining optical counterparts to unknown EUV sources in the ROSAT WFC
catalogues. In the absence of any plausible EUV source, 
the 15th magnitude G-type central star 
in the error box may be hiding a hot white dwarf companion. The far-UV
spectra obtained with IUE (SWP56272 and LWP31800) were very
noisy and showed no evidence for a hot white dwarf. There was some flux
above the background longwards of $\sim$2700{\AA} in the LWP spectrum,
which may have been due to a G star, 
but it is  possible that the
target was missed altogether, and this flux was due to scattered solar
light which effects the LWP camera sporadically. 
At V$=$14.8 this star is unlikely to be the source of the EUV flux: a
15th magnitude main sequence mid-G star would require
L$_{EUV}$/L$_{bol}$$\sim$0.1 to produce the count rate seen in the WFC S2
filter (45 counts ksec$^{-1}$), far in excess of the saturation level for
coronal emission (L$_{EUV}$/L$_{bol}$$\sim$10$^{-3}$, Mathioudakis et al.
1995).

\subsubsection{SAO150508 (2RE J0530$-$19)}

Prior to the publication of the 2RE catalogue (Pye et al. 1995), the 9th
magnitude F6V star SAO150508 was not thought to be active, although there are no
published optical observations which might offer evidence one way or
another. The IUE spectra (SWP 56195 and LWP31700, Figure 7) 
show no evidence for a hot white dwarf companion. 
SAO150508 is, however, coincident with a ROSAT PSPC X-ray source.
Therefore, estimates of the X-ray and EUV luminosities, 
assuming SAO150508 is active and  
the true source of the EUV and X-ray flux, are presented in Table 7.

\subsubsection{HD36869 (2RE J0534$-$15)}

There is no evidence in the literature that the 8th magnitude G2V star 
HD36869 is active. The IUE spectra (SWP56169 and LWP 31701, Figure 8)
also show no evidence for activity, although the star is coincident with a ROSAT
PSPC source. Given that 8th magnitude stars are comparatively rare, it is
still possible that HD36869 is coronally active, and thus we provide estimates
of the X-ray and EUV luminosities in Table 7.

\subsubsection{Gl216B (2RE J0544$-$22)}

Gl216B (K2V) is part of a nearby ($\approx$8 parsecs) triple system, and
was chosen as a candidate white dwarf binary on the basis of the
S2/S1 count rate ratio. The IUE 
spectra (SWP56194 \& LWP31699, Figure 9) show no evidence for a hot white dwarf.
However, Mg II 2798{\AA} is visible in emission (with a line flux of
4.2$\pm$0.7$\times$10$^{-12}$ ergs cm$^{-2}$ s$^{-1}$ above the
continuum). The emission feature in the SWP spectrum at
$\sim$1720{\AA} is probably spurious, since it does not coincide with any
commonly seen line. From observations made in the optical, de Strobel et al. (1989) 
concluded that this is a young, active star. If it is the only source of
the EUV flux, then we determine
L$_{EUV}$/L$_{bol}$$=$2.78$\times$10$^{-5}$. 

Active stars have a characteristic EUV to Mg II flux ratio. For example,
Jewell (1993) shows that the 0.05$-$0.2 keV EUV flux is 1$-$10 $\times$
the MgII flux. For Gl216B, though, the EUV to Mg II flux ratio is only
$\approx$0.8. 

Schmitt et al. (1990) observed the
entire Gl216 system in an Einstein HRI pointing, and found that the nearby F7V star
Gl216A was 8 times brighter in X-rays than Gl216B. Thus
Hodgkin and Pye (1994) concluded that all of the EUV radiation in fact comes from
Gl216A. However, we only targeted Gl216B with IUE since this is the object  
associated with the EUV source in the 2RE catalogue, and at the time of the 
observation we were unaware of Hodgkin and Pye's conclusion. The low EUV to
Mg II flux ratio for Gl216B does, though, support these earlier conclusions
that the major source of the EUV radiation is actually Gl216A.

Estimates of the X-ray and EUV luminosities are given in Table
7 assuming a) all the flux comes from Gl216A and b) all the flux comes
from Gl216B. 

\subsubsection{HR2225 (HD43162, RE J0613$-$23)}

This G5V star was not known to be active prior to the ROSAT survey, but
subsequently it has been studied in detail by Jeffries and Jewell (1993) and is almost
certainly the EUV source. It is also an X-ray source, and measurements of
the X-ray and EUV luminosities are given in Table 7. No obvious emission
features are visible in the IUE LWP and SWP spectra (Figure 10); 
the feature 
longwards of 1800{\AA} in the SWP spectrum is probably spurious, 
as there is no commonly seen line at this wavelength. 

\subsubsection {HD295290 (2RE J0640$-$03)}

There are no references in the literature to this being an active star.
However, Mg II is clearly seen in emission at 2800{\AA} in the IUE LWP
spectrum (Figure 11), and there is a suggestion of CIV in emission at
1550{\AA} in the short wavelength region. Measurements of the X-ray and
EUV luminosities are given in Table 7 using the G0 classification
given by SIMBAD, and assuming the star is on the main sequence. The EUV
to Mg II flux ratio ($\approx$8.0) strongly suggests that this star is
active and the true source of the EUV radiation. Note that
the ratios L$_{EUV}$/L$_{bol}$ and L$_{x}$/L$_{bol}$ are significantly
larger than for any of the other stars in this sample, approaching the
saturated level for coronal emission ($\sim$10$^{-3}$). This suggests that
the star may be rapidly rotating. 

\subsubsection {HD54402, (2RE J0710$+$45)}

No references are given in the literature to this being an active star.
There is clearly no white dwarf visible in the IUE SWP spectrum
(Figure 12), and the
emission line at $\sim$1800{\AA} is probably spurious, perhaps due to a
cosmic ray hit. The EUV source is not coincident with an X-ray source.
The star needs to be examined optically to search for any evidence of
chromospheric activity.

\subsubsection {SAO135659, (2RE J0813$-$07)}

Again, this star was not known to be active prior to the ROSAT survey.
The IUE LWP spectrum (Figure 13) reveals Mg II in emission at 2800{\AA}.
Measurements of the EUV and X-ray luminosities (this star is also a PSPC
source) are given in Table 7, assuming the  K0 spectral type given by
SIMBAD, and that the star is on the main sequence. The EUV
to Mg II flux ratio ($\approx$8.25) strongly suggests that this star is
active and the true source of the EUV radiation.

\subsubsection {RE J0823$-$252/HD70907}

With S1 and S2 count rates of 52$\pm$7 and 83$\pm$9 counts ksec$^{-1}$, this is a
relatively bright EUV source in comparison with most of the targets in this paper. 
The soft X-ray and EUV photometric colours are also 
characteristic of a hot white dwarf, and it is not detected in the PSPC hard
band. Therefore,  it was selected as a potential hidden white dwarf binary. 

The V$=$8.8 star in the centre of the field,
HD70907 (F3IV/V), was observed in both the IUE SWP and LWP cameras
(Figure 14).
There is no evidence for a white dwarf companion or  emission features
indicative of an active star. A nearby V$\approx$11 star was also observed
and again there was no evidence for a white dwarf, although,  in
the absence of any flux in the LWP camera 
that could be attributed to a stellar source, 
it seems possible that the star was not in the LWP slit.  
Mason et al. (1995) report that this fainter object is indeed active, although
they give no indication of the size of any emission features seen in the
optical. They also do 
not give a spectral type for this star, and thus we have not been able to
determine the EUV and X-ray luminosities. Whether this star is active
enough to be the true EUV source remains unclear, and the suspicion remains that there
is indeed an unresolved hot white dwarf hiding in this field. 

\subsubsection{HR4646 (2RE J1212$+$77)}

It is highly unusual to detect an A star in EUV or X-ray surveys
(Fleming et al. 1991).
Observations by the Einstein and EXOSAT observatories failed to find any
convincing detections other than the nearby quadruple A star system
Castor (Pallavicini et al. 1990). Therefore HR4646, an Am star coincident
with ROSAT WFC 2RE and PSPC sources, was selected as a potential hidden
white dwarf binary. It should be noted that Am stars do not possess
significant magnetic fields and  they are slow rotators, but they
almost always appear to lie in close binary systems (Abt 1961), and
indeed Margoni, Munari and Stagni (1992)  
found that HR4646 is a spectroscopic binary with a period of 1.27 days. 

The IUE SWP spectrum (Figure 15) shows no evidence
for a hot white dwarf companion. It is extremely likely, therefore, that 
HR4646 has a coronally active cooler companion (F5 or later).

\section{Summary and Conclusions}

A search for unresolved white dwarfs in binary systems with optically brighter
normal stellar companions has been conducted with IUE during its final
year of operation. Targets were
chosen among the fainter and still un-identified EUV sources in the ROSAT
2RE catalogue (Pye et al. 1995). One new system was discovered (RE
J0500$-$364, DA$+$F6/7V), which appears to lie in a direction of low
interstellar neutral hydrogen volume density. 
If this star is really as distant as  $\sim$500-1000pc, then it may
represent a lower limit to the size of the $\beta$~CMa tunnel of low
density neutral gas stretching away from the Local Bubble in that
direction. 

Including the independently identified HD27483 (DA$+$F6V, B\"ohm-Vitense
1993), this new discovery 
brings the total number of such systems found in the ROSAT and EUVE
surveys to nineteen. These two satellites have, therefore, been very
successful in helping us to identify these kinds of binary systems, which have
never been seen optically and could only have been identified through such
satellite surveys. However, these stars still represent $<$20\% of the hot 
white dwarfs identified in the EUV.  

The demise of IUE, which suffered a gyro failure in February 1996,
limiting it to observe only targets with bright guide stars (the mission was
finally terminated in September 1996), effectively  
cut short our programme. The loss of IUE means that this particular
method of searching for these important systems is no longer available.
Although HST can observe the same wavelength region, it is of course much
harder to obtain the time required for this kind of search. It is likely,
therefore, that very few new examples of these systems will be discovered
in the near future. The results presented in this paper seem to suggest
that this search, using the EUV catalogues as a basis from which 
to identify potential systems, has been fairly exhaustive (we
have identified only one new system from thirteen targets). 

However, the identification by Vennes et al. (1997) 
of a hot white dwarf companion to a previously catalogued EUV-bright active 
star, RE J0702$+$129 (K0Ve), instead suggests that, in fact, some systems 
may have been completely overlooked. RE J0702$+$129 was classified as active by
Mason et al. (1995) in their optical follow-up programme to identify the
EUV sources found in the  ROSAT WFC survey, and was not, therefore,
included on any of our target lists for the IUE search. This raises the
question of how many of the $>$200 `active' stars in the WFC and EUVE
catalogues really are the source, or at least the only source, of the EUV
radiation. Until each star has been observed and analysed in detail, the
suspicion remains that there are more of these binary systems in the survey
waiting to be identified. 

For example, 
the ROSAT WFC catalogue includes $\approx$120 isolated white dwarfs, and
$\approx$60 in some kind of binary, e.g. classic Sirius-type systems, CVs,
non-interacting DA$+$dM systems, and visual binaries. Conservatively,
then, we might assume that we have already identified the majority of the white
dwarf binaries to be found in the survey. However, if as many as 80\% of stars
reside in binary or multiple systems, another 30 might be awaiting
discovery. These could include double degenerate systems as well as further
examples of Sirius-type systems (e.g. 
the apparantly isolated DA RE J0512$-$007 has a mass, M$=$0.38M$_\odot$, 
too low for it to be the result of single star evolution, 
and may have a degenerate companion).  

What is really needed to try to find more Sirius-type binaries
 is an all-sky UV survey. None has been undertaken since the TD-1
survey of 1972/73, which originally appeared to have 
only detected Sirius B among these systems.  
Subsequently, though, Landsman, Simon and Bergeron (1996) found white
dwarf companions to 56 Persei and HR3643 as a result of the UV excess
detected by TD-1. The recently approved \emph{GALEX} mission (Bianchi
1998), which will
survey the sky at UV wavelengths and follow-up some targets
spectroscopically, may reveal many more of these binaries.  

In the meantime, follow-up observations of these 
nineteen new EUV-bright systems are
required. In particular, it is important to determine whether these
systems are wide, or close enough that they must have undergone Common
Envelope evolution. This information will help to place constraints on
theoretical models of binary evolution (e.g. de Kool and Ritter 1993).
Detailed studies of the normal stellar companions may also reveal
evidence of past interaction (e.g. Jeffries and Stevens 1996), and stars
with possible abundance anomalies that may be the progenitors of the
Barium and Carbon giants (e.g. Jeffries and Smalley 1996). 

In addition, the forthcoming \emph{FUSE} mission (Far Ultra-violet 
Spectroscopic Explorer) will, for the first time, allow us to unambiguously 
determine the temperatures and surface gravities of the white dwarfs in these
systems (and hence their masses and radii), through modelling of the H
Lyman absorption series down to 912{\AA}.

\section*{Acknowledgements}MRB and MAB acknowledge the support of PPARC, UK. 
We wish to thank Detlev Koester (Kiel) for the use of his white
dwarf model atmosphere grids, and Jim Collins (Arizona) and 
Nigel Bannister (Leicester) for their help 
in obtaining and reducing the optical spectrum of RE J0500$-$364. 
MRB wishes to thank the staff at IUE Vilspa
for their help and co-operation in carrying out this programme. 
We also thank the referee, Rob Jeffries, for useful suggestions to
improve this paper. This research has made use of the
SIMBAD database operated at CDS, Strasbourg, France, and data obtained from
the Leicester Database and Archive Service (LEDAS) at
the Department of Physics and Astronomy, Leicester University, UK.

\newpage

\section*{Figure Captions}

Figure 1: Field of the ROSAT WFC EUV source RE J0500$-$364 (6$\times$6
arcmins). The circles
are the 90\% error boxes, centred on the source co-ordinates, of (in
order of decreasing size) EUVE, the WFC, and the ROSAT HRI. The WFC error
box is centred on co-ordinates (J2000) 05 00 03.9 $-$36 24 02. The DA+F6/7V
binary lies near the centre of each error box. The arrows
indicate  other stars examined by Mason et al. (1995) and ourselves
during optical searches for the EUV and X-ray counterpart. None of these
objects was particularly remarkable. \\

Figure 2: Optical spectrum of RE J0500$-$364 obtained with the Steward
Observatory 2.3m telescope on Kitt Peak. Using this spectrum, we classify
the star as F6/7V.\\

Figure 3: Far-ultraviolet IUE SWP spectrum of RE J0500$-$364 (SWP56217),
obtained in December 1995 (8 hour exposure), clearly showing the white
dwarf companion. Also shown is a 
pure H model atmosphere fit for log g$=$7.25, T$=$38,260K and
M$=$0.40M$_\odot$.\\

Figure 4: Far-ultraviolet IUE LWP spectrum of RE J0500$-$364 (LWP31729).
The data are very noisy and have been binned up, but the spectral shape
most closely matches stars in the range F5V$-$G0V, consistent with the
optical data.\\

Figure 5: Pulse height distribution of the X-rays from RE J0500$-$364  
detected by the ROSAT HRI. Although the exact energies of each channel
are known to vary spatially and temporally, it can be seen that this is a
soft source, and there are no hard counts. \\

Figure 6:  Far-ultraviolet IUE SWP spectrum of HD27483 (SWP45940). The
hot white dwarf can clearly be seen emerging from the glare of
the two F6V companions shortwards of $\sim$1700{\AA}. Also shown is a
pure H model atmosphere fit for log g$=$8.5, T$=$22,000K and
M$=$0.94M$_\odot$.\\

Figure 7: Far-ultraviolet IUE spectrum of SAO150508
(G5V, SWP56195$+$LWP31700). Clearly, there is no white dwarf companion to
this
star. \\

Figure 8: Far-ultraviolet IUE spectrum of HD36869 (G2V,
SWP56169$+$LWP31701). No white dwarf companion is visible in the short
wavelength spectrum. \\ 

Figure 9: Far-ultraviolet IUE spectrum of Gl216B (K2V,
SWP56194$+$LWP31699). Mg II 2798{\AA} is visible in emission. The feature
at $\sim$1720{\AA} (inset) is probably due to a cosmic ray hit. \\

Figure 10: Far-ultraviolet IUE spectrum of HR2225 (G5V,
SWP56206$+$LWP31715). The emission feature at $\sim$1800{\AA} (inset) is
probably spurious. \\

Figure 11: Far-ultraviolet IUE spectrum of HD295290 (G0V,
SWP56211$+$LWP31726). MgII is clearly seen in emission at 2798{\AA}. CIV
1549{\AA} emission is also suggested in the SWP spectrum (inset). \\

Figure 12: Far-ultraviolet IUE spectrum of HD54402 (K0,
SWP56193$+$LWP31698). The feature at $\sim$1800{\AA} is probably due to a
cosmic ray hit. \\

Figure 13: Far-ultraviolet IUE spectrum of SAO135659
(K0, SWP56273$+$LWP31801). MgII is visible in emission at 2798{\AA}. \\

Figure 14: Far-ultraviolet IUE spectrum of HD70907 (F3IV/V,
SWP56344$+$LWP31836).
There is no evidence for activity. \\

Figure 15: Far-ultraviolet IUE SWP spectrum of HR4646 (A5m, SWP55658).
Clearly, there is no white dwarf companion to this star. \\

\newpage

\begin{table*}
\small 
\begin{center}
\caption{ROSAT WFC/PSPC and EUVE count rates (1000s$^{-1}$)}
\begin{tabular}{llcccccc}

 &  & WFC & & PSPC & & EUVE & \\
ROSAT No. & Name$^\ddag$ & S1 & S2 & (0.1$-$0.4keV) & (0.4$-$2.4keV) & 100\AA
& 200\AA \\

WD Binaries & & & & & & & \\
RE J0500$-$36$^{a}$ & & 20$\pm$5 & 47$\pm$8 & no det. & no det. &
21$\pm$6 & no det. \\
RE J0420$+$13$^{b}$ & HD27483 & 15$\pm$4 & 27$^{\dag}$ & 98$\pm$14 &
33$\pm$9 & no det. & no det. \\
Non-WD Systems & & & & & & \\
2RE J0222$+$50 & BD$+$49$^\circ$646 & 14$\pm$3 & 17$\pm5$ & 258$\pm$20 & 
195$\pm$18 & no det. & no det. \\
2RE J0232$-$02 & & 15$^{\dag}$ & 45$\pm$11 & no det. & no det. & no det. &
no det. \\
2RE J0530$-$19 & SAO150508 & 8$\pm$3 & 16$\pm$5 & 159$\pm$18 &
195$\pm$20 & no det. & no det. \\
2RE J0534$-$15 & HD36869 & 8$\pm$3 & 19$\pm$5 & 221$\pm$22 & 217$\pm$22
& no det. & no det. \\
2RE J0544$-$22 & Gl 216B & 7$\pm$3 & 22$\pm$5 & 261$\pm$23 & 82$\pm$13 &
no det. & no det. \\
RE J0613$-$23 & HR2225 & 18$\pm$3 & 35$\pm$9 & 517$\pm$30 & 256$\pm$21 &
no det. & no det. \\ 
2RE J0640$-$03 & HD295290 & 9$\pm$3 & 32$\pm$8 & 304$\pm$28 & 281$\pm$27
& no det. & no det. \\
2RE J0710$+$45 & HD54402 & 8$\pm$3 & 26$\pm$7 & no det. & no det. & no
det. & no det. \\
2RE J0813$-$07 & SAO135659 & 14$\pm$6 &  55$\pm$16 & 207$\pm$35 &
195$\pm$34 & no det. & no det. \\
RE J0823$-$25$^{c}$ & & 52$\pm$7 & 83$\pm$9 & 175$\pm$24 & no det. &
28$\pm$11 & no det. \\
2RE J1212$+$77 & HR4646 & 13$^{\dag}$ & 18$\pm$5 & 157$\pm$42 & 114$\pm$36
& no det. & no det. \\

\end {tabular}
\end{center}
$\dag$ upper limit \\
a] Detected in the First EUVE Source Catalog only (Bowyer et al., 1994) \\
b] ROSAT WFC Bright Source Catalogue detection only (Pounds et al., 1993) \\ 
c] HD70907 and a V$\approx$11 companion were both observed as possible
counterparts to this source    \\
\normalsize
\end{table*}

\begin{table*}
\begin{center}
\caption{Log of IUE observations}
\begin{tabular}{lcccccl}

Name & SWP No. & LWP No. & Date & Exp. (s) & Observer & notes \\

WD Binaries & & & & & & \\
RE J0500$-$364 & 56217 & & 1995/323 & 1800 & SO &  \\
 & 56338 & & 1995/358 & 29400 & SO & 159DN problem \\
 & & 31729 & 1995/323 & 1800 & SO & \\ 
HD27483 & 45940 & & 1992/273 & 2100 & B\"ohm-Vitense & \\ 
Non-WD systems & & & & & & \\
2RE J0222$+$50 & 56333 & & 1995/358 & 1800 & SO &  \\
2RE J0232$-$02 & 56272 &  & 1995/340 & 1800 & SO &  \\
 & & 31800 & 1995/340 & 1800 & SO &  \\
SAO150508 & 56195 & & 1995/317 & 1800 & SO &  \\
 & & 31700 & 1995/317 & 1200 & SO &  \\
HD36869 & 56196 & & 1995/317 & 1800 & SO &  \\
 & & 31701 & 1995/317 & 360 & SO &  \\
Gl216B & 56194 & & 1995/317 & 1800 & SO &  \\
 & & 31699 & 1995/317 & 120 & SO &  \\
HR2225 & 56206 & & 1995/319 & 1800 & SO &  \\
 & & 31715 & 1995/319 & 60 & SO &  \\ 
HD295290 & 56211 & & 1995/322 & 1800 & SO &  \\
 & & 31726 & 1995/322 & 1200 & SO & 159DN problem \\
HD54402 & 56193 & & 1995/317 & 1800 & SO &  \\
 & & 31698 & 1995/317 & 300 & SO &  \\
SAO135659 & 56273 & & 1995/340 & 1800 & SO &  \\
 & & 31801 & 1995/340 & 1200 & SO &  \\
HD70907 & 56344 & & 1995/359 & 1800 & SO &  \\
 & & 31836 & 1995/359 & 300 & SO &  \\
RE J0823$-$25$^{\dag}$ & 56266 & & 1995/338 & 1800 & SO &  \\
 & & 31796 & 1995/338 & 300 & SO &  \\
HR4646 & 56393 & & 1996/009 & 100 & SO &  \\

\end {tabular}
\end{center}
SO$=$ Service Observation \\ 
$\dag$ V$\approx$11 companion to HD70907 \\ 
\end{table*}

\begin{table*}
\begin{center}
\caption{Physical parameters of the companion stars in the WD binaries}
\begin{tabular}{lcccccl}

RE No. & Cat. Name & SpT &  V  & d est. (pc) & references \\

RE J0500$-$36 & & F6/7V & 13.7 & 520$-$1250 &  \\
RE J0420$+$13 & HD27483 & F6V & 6.17  & 47.6 & B\"ohm-Vitense (1993) \\

\end {tabular}
\end{center}
\end{table*}

\begin{table*}
\begin{center}
\caption{Physical parameters of the stars observed where no white
dwarf was detected}
\begin{tabular}{lcccccl}

RE No. & Cat. Name & Spectral type &  V magnitude &
references \\

2RE J0222$+$50 & BD$+$49$^\circ$646 & ? & 10.1 & SIMBAD \\
2RE J0232$-$02 & & K & 14.8 & SIMBAD \\
2RE J0530$-$19 & SAO150508 & G5V & 9.0 & SIMBAD \\
2RE J0534$-$15 & HD36869 & G2V & 8.0 & SIMBAD \\
2RE J0544$-$22 & Gl216B & K2V & 6.15 & SIMBAD \\
RE J0613$-$23 & HR2225 & G5V & 6.39 & SIMBAD \\ 
2RE J0640$-$03 & HD295290 & G0 & 9.1 & SIMBAD \\
2RE J0710$+$45 & HD54402 & K0 & 7.7 & SIMBAD \\
2RE J0813$-$07 & SAO135659 & K0 & 8.8 & SIMBAD \\
RE J0823$-$25 & HD70907 & F3IV/V & 8.8 & SIMBAD \\
  &  & late & 11 & Mason et al. (1995) \\
2RE J1212$+$77 & HR4646 & A5m & 5.14 & SIMBAD \\

\end {tabular}
\end{center}
\end{table*}

\begin{table*}
\small
\begin{center}
\caption{Temperatures and gravities for the white dwarfs from
homogeneous model fits}
\begin{tabular}{lccccccc}

Binary & log g &  Temp & 90\% error & Mass & Radius & d $\sb {wd}$ &
Estimated V \\
     &  & K & K & M$_\odot$ & R$_\odot$ & (pc) & \\

RE J0500$-$36 & 7.0 & 36,800 & 33,900-41,000 & 0.38 & 0.032 & 1050 & 17.8 \\
 & 7.25 & 38,260 & 35,350-42,900 & 0.40 & 
0.025 & 830 & 17.9 \\
 & 7.5 & 42,000 & 36,400-45,700 &  0.47 & 
 0.020 &  740 &  17.9 \\
 & 8.0 & 45,000 & 40,500-50,500 & 0.68 & 0.014 & 520 & 17.9 \\
 & 8.5 & 50,500 & 45,400-57,700 & 0.97 & 0.009 & 390 & 18.0 \\
 & 9.0 & 60,500 & 51,900-67,500 & 1.21 & 0.006 & 270 & 18.1 \\
HD27483 & 7.0 & 20,000 & upper limit & 0.37 & 0.032 & 103 & 13.9 \\
 & 7.5 & 20,000 & upper limit & 0.40 & 0.019 & 61 & 13.9 \\
 &  8.0 & 21,410 & 21,215-21,630 & 0.63 & 0.013 & 50 & 14.2 \\
 & 8.5 & 22,000 &  21,910-22,780 & 0.94 & 
 0.009 & 44 &  14.5 \\
 & 9.0 & 24,800 & 24,570-25,060 & 1.20 & 0.006 & 30 & 14.6 \\

\end {tabular}
\end{center}
\normalsize
\end{table*}

\begin{table*}
\small
\begin{center}
\caption{Column densities from homogeneous models for RE
J0500$-$362 (assuming pure H composition)}
\begin{tabular}{lccccl}

Name & log g & T & HI column & 90\% error & Comment \\

RE J0500$-$364 & 7.0 & 36,800 &
3.7$\times$10$^{18}$ & 1.7$-$7.0$\times$10$^{18}$ & \\
 & 7.25 & 38,260 & 7.5$\times$10$^{18}$ & 
 5.0$\times$10$^{18}$$-$1.1$\times$10$^{19}$  & \\
 & 7.5 & 42,000 & 1.7$\times$10$^{19}$ & 1.4$-$2.1$\times$10$^{19}$ & \\
 & 8.0 & 45,000 & 2.4$\times$10$^{19}$ & 2.1$-$2.9$\times$10$^{19}$ & \\
 & 8.5 & 50,500 & - & - & No fit \\
 & 9.0 & 60,500 & - & - & No fit\\

\end {tabular}
\end{center}
No fit could be obtained to the ROSAT data for HD27483 \\
\normalsize
\end{table*}

\begin{table*}
\begin{center}
\caption{EUV and X-ray luminosities for the probable and 
confirmed active stars}
\begin{tabular}{lccccccc}

 Name & d(pc) & L$_{EUV}$ & L$_{EUV}$/L$_{bol}$ & L$_x$ &
L$_x$/L$_{bol}$ & f$_{MgII}$ & L$_{MgII}$/L$_{bol}$ \\
 & & $\times$10$^{29}$ ergs & $\times$10$^{-4}$ & $\times$10$^{29}$
ergs & $\times$10$^{-4}$ & $\times$10$^{-12}$ & $\times$10$^{-4}$ \\

 SAO150508 & 59 & 10.0 & 3.6 & 1.3 & 4.7 & - & - \\
 HD36869 & 48 & 7.9 & 1.7 & 10.0 & 2.2 & - & - \\
 Gl216A & 8 & 0.3 & 0.04 & 0.2 & 0.02 & - & - \\
 Gl216B & 8  & 0.3 & 0.3 & 0.2 & 0.2 & 4.2 & 0.5  \\
 HR2225 & 18 & 2.0 & 0.7 & 2.0 & 0.7 & - & - \\
 HD295290 & 95 & 51.8 & 8.2 & 51.5 & 8.2 & 0.6 & 1.0 \\
 SAO135659 & 34 & 11.4 & 0.7 & 4.5 & 0.4 & 1.0 & 1.1 \\

\end{tabular}
\end{center}
\end{table*}

\begin{table*}
\begin{center}
\caption{Neutral hydrogen column and volume densities in the local
interstellar medium}
\begin{tabular}{lcccccl}

 Name & l & b & d(pc) & N$_H$ (cm$^{-2}$) & nH (cm$^{-3}$) & Ref. \\

 $\beta$~CMa & 226.1 & $-$14.3  & 206 & 2.2$\times$10$^{18}$$^{\dag}$ & 0.0035 &
Cassinelli et al. 1995 \\
 $\epsilon$CMa & 239.8 & $-$11.3 & 188  & 1.2$\times$10$^{18}$$^{\dag}$ &
0.0021 & Cassinelli et al. 1996 \\
 RE J0457$-$281 & 229.3 & $-$36.2  & 90 & 9.6$\times$10$^{17}$ & 0.0035 &
Barstow 1997 {\it private com.} \\
 RE J0503$-$289 & 230.7 & $-$34.9 & 90 & 8.2$\times$10$^{17}$ & 0.0030 &
Barstow 1997 {\it private com.} \\
 RE J0500$-$364 & 239.6 & $-$36.2 & 830 & 7.5$\times$10$^{18}$ & 0.0029 &
this paper \\

$^{\dag}$ Upper limit \\
\end{tabular}
\end{center}
\end{table*}


\begin{thebibliography}{99}

\bibitem{} Abt H.A., 1961, ApJS, 6, 37

\bibitem{} Allen C., 1973, {\it Astrophysical Quantities}, Athlone Press,
London

\bibitem{} Barstow M.A., Schmitt J.H.M.M., Clemens J.C., Pye J.P., Denby
M., Harris A.W., Pankiewicz G.S., 1992, MNRAS, 255, 369

\bibitem{} Barstow M.A. et al., 1993, AdSpR, 131, 281

\bibitem{} Barstow M.A. et al., 1994, MNRAS, 270, 499

\bibitem{} Bianchi L., 1998, in: Ultraviolet astrophysics
$-$ Beyond the IUE final archive, ESA publication SP$-$413, 797

\bibitem{} B\"ohm-Vitense E., 1980, ApJ, 239, L79 

\bibitem{} B\"ohm-Vitense E., 1993, Astron J., 106, 1113

\bibitem{} B\"ohm-Vitense E., 1995, AJ, 110, 228

\bibitem{} Bohlin R.C., Grillmair C.J., 1988, ApJS, 66, 209

\bibitem{} Bowyer S. et al., 1994, ApJS 93, 569

\bibitem{} Bowyer S. et al., 1996, ApJS, 102, 129

\bibitem{} Burleigh M.R., Barstow M.A., 1997, in: White Dwarfs, eds. J.
Isern, M. Hernanz and E. Garcia-Berro, Kluwer, 329

\bibitem{} Burleigh M.R., Barstow M.A., 1998, MNRAS, 295, L15

\bibitem{} Burleigh M.R., Barstow M.A., Fleming T.A., 1997, MNRAS, 287,
381 (Paper I)

\bibitem{} Cassinelli J.P. et al., 1995, ApJ, 438, 932

\bibitem{} Cassinelli J.P. et al., 1996, ApJ, 460, 949

\bibitem{} Dupuis J., Vennes S., Chayer P., Cully S., Rodriguez-Bell T.,
1997, in: White Dwarfs, eds. J. Isern, M. Hernanz
and E. Garcia-Berro, Kluwer, 277

\bibitem{} ESA, 1997, {\it The Hipparcos Catalogue}, ESA SP$-$1200

\bibitem{} Fleming T.A., Liebert J., Green R.G., 1986, ApJ, 308, 176

\bibitem{} Fleming T.A., Schmitt J.H.M.M., Barstow M.A., Mittaz J.P.D., 
1991, A\&A, 246, L47

\bibitem{} Fleming T.A. et al., 1995, ApJS, 99, 701 

\bibitem{} Fleming T.A., Snowden S.L., Pfefferman E., Briel U., Greiner
J., 1996, A\&A, 316, 147

\bibitem{} Frisch P.C., York D.G., 1983, ApJ, 271, L59

\bibitem{} Garhart M., 1997, IUE Newsletter (GSFC) Vol. 5, No. 5

\bibitem{} Hodgkin S.T., Barstow M.A., Fleming T.A., Monier R., Pye J.P., 
1993, MNRAS, 263, 229

\bibitem{} Hodgkin S.T., Pye J.P., 1994, MNRAS, 267, 840

\bibitem{} Koester D., 1991, IAU Symposium 145, Evolution of Stars: The
Photospheric Abundance Connection, eds. G. Michaud and A. Tutukov, Kluwer
Dordrecht, 435

\bibitem{} de Kool M., Ritter H., 1993, A\&A, 267, 397

\bibitem{} Jacoby J.H., Hunter D.A., Christian C.A., 1984, ApJS, 56, 257

\bibitem{} Jeffries R.D., Jewell S.J., 1993, MNRAS, 264, 106

\bibitem{} Jeffries R.D., Burleigh M.R., Robb R.M., 1996, A\&A, 305, L45 

\bibitem{} Jeffries R.D., Stevens I.R., 1996, MNRAS, 279, 180

\bibitem{} Jeffries R.D., Smalley B., 1996, A\&A, 315, L19

\bibitem{} Jewell S.J., 1993, PhD Thesis, University of Birmingham

\bibitem{} Johnson H.R., Ake T.B., 1986, ESA SP-263, 395

\bibitem{} Jordan, S., 1997, in: White Dwarfs, eds. J. Isern, M. Hernanz
and E. Garcia-Berro, Kluwer, 397

\bibitem{} Landsman W., Simon T., Bergeron P., 1996, PASP, 108, 250

\bibitem{} Margoni R., Munari U., Stagni R., 1992, A\&AS, 93, 545

\bibitem{} Marsh M.C. et al., 1997, MNRAS, 286, 369

\bibitem{} Mason K.O. et al., 1995, MNRAS, 274, 1194

\bibitem{} Mathioudakis M., Fruscione A., Drake J.J., McDonald K., Bowyer
S., Malina R.F., A\&A, 300, 775

\bibitem{} McCook G.P. and Sion E.M., 1998, ApJS, in preparation

\bibitem{} Pallavicini R., Tagliaferri G., Pollock A.M.T., Schmitt
J.H.M.M., Rosso C., 1990, A\&A, 227, 483

\bibitem{} Pounds K.A. et al., 1993, MNRAS, 260, 77

\bibitem{} Pye J.P. et al., 1995, MNRAS, 274, 1165

\bibitem{} Schindler M., Stencel R.E., Linsky J.L., Basri G.S., Helfand
D.J., 1982, ApJ, 263, 269

\bibitem{} Schafer R.A. et al., 1991, ESA TM-09

\bibitem{} Schaller G., Schaerer D., Meynet G., Maeder A., 1992, A\&AS,
96, 269

\bibitem{} Schmitt J.H.M.M., Collura A., Sciortino S., Vaiana G.S.,
Harnden F.R., Rosner R., 1990, ApJ, 365, 704

\bibitem{} Shipman H.L., Geczi J., in `White Dwarfs', ed. G. Wegner,
Springer-Verlag, 134

\bibitem{} Vennes S., Mathioudakis M., Doyle J.G., Thorstensen J.R.,
Byrne P.B., 1995, A\&A, 299, L29 

\bibitem{} Vennes S., Berghoefer T., Christian D.J., 1997, ApJ, 491, L85

\bibitem{} Vennes S., Christian D.J., Mathioudakis M., Doyle J.G., 1997,
A\&A, 318, L9

\bibitem{} Vennes S., Thejll P.A., Genova Galvan R., Dupuis J., 1997b,
ApJ, 480, 714

\bibitem{} Voges W., et al., 1996, IAU Circ. 6420

\bibitem{} Warwick R.S., Barber C.R., Hodgkin S.T., Pye J.P., 1993,
MNRAS, 262, 289

\bibitem{} Welsh, B.Y., 1991, ApJ, 373, 556

\bibitem{} Wesemael F., Auer L.H., van Horn H.M., Savedoff M.P., 1980,
ApJS, 43, 159

\bibitem{} Wood M.A., 1992, ApJ, 386, 539

\bibitem{} Wood M.A., 1995, in Proc. of the 9th European Workshop on
White Dwarfs, eds. D. Koester and K. Werner, Springer, 41

\bibitem{} Wu C.-C. et al., 1992, {\IUE} Ultraviolet Spectral Atlas of
Selected Astronomical Objects, NASA Reference Publication 1285

\end{thebibliography}
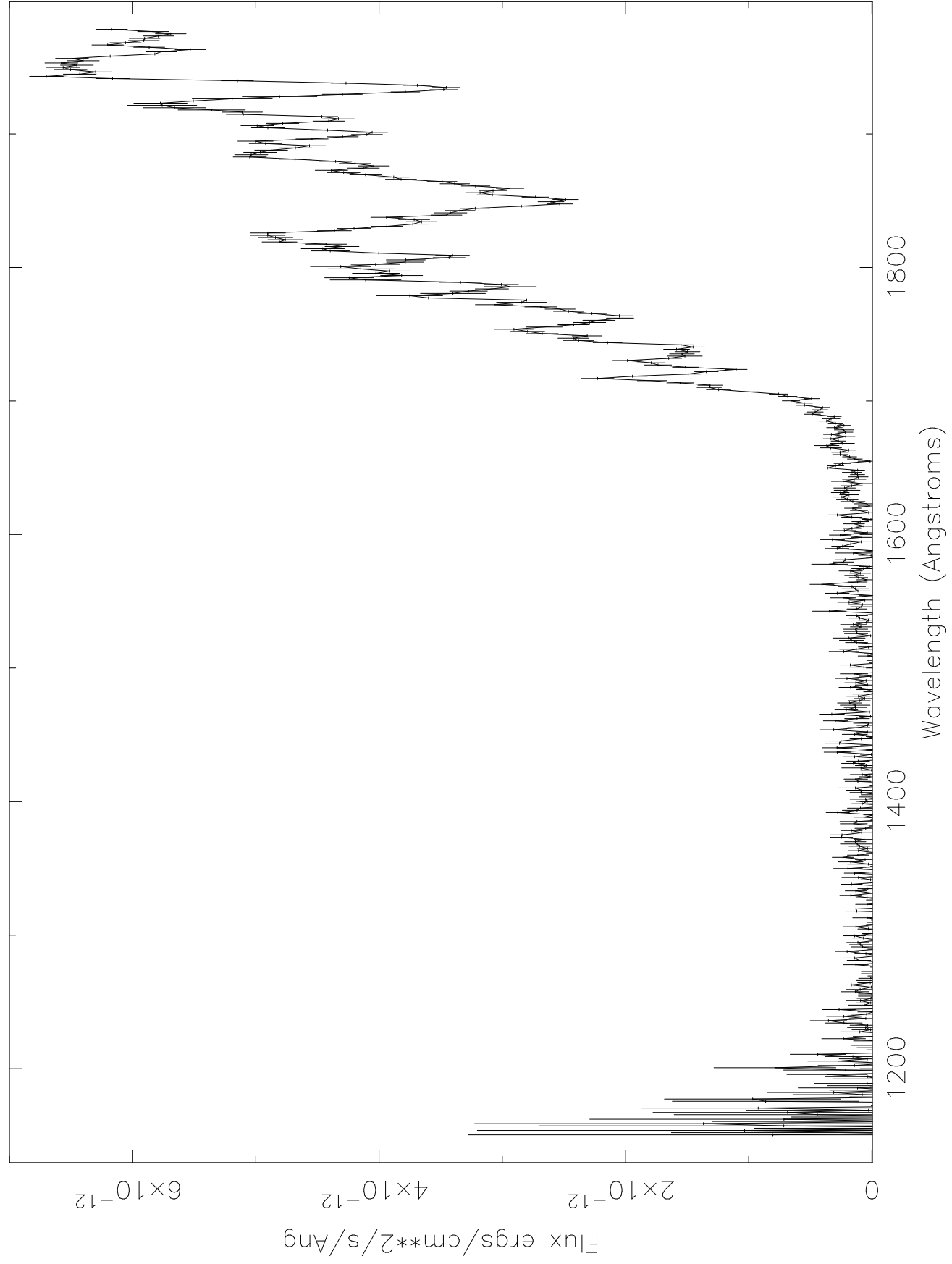
\end{document}